\documentclass[10pt]{article}
\usepackage{fullpage}
\usepackage{amsmath}
\usepackage{amssymb}
\usepackage{amsfonts}
\usepackage[dvips]{epsfig}
\usepackage{color}

\def\##1{{\bf #1}}
\def\=#1{\underline{\underline{#1}}}

\def\+#1{\underline{\bf #1}}
\def\*#1{\underline{\underline{\bf #1}}}

\def\r#1{(\ref{#1})}
\def\l#1{\label{#1}}
\def\c#1{\cite{#1}}

\def\le{\left(}
\def\ri{\right)}
\def\les{\left[}
\def\ris{\right]}
\def\lec{\left\{}
\def\ric{\right\}}

\def\.{\cdot}

\def\epso{\epsilon_{\scriptscriptstyle 0}}

\def\muo{\mu_{\scriptscriptstyle 0}}

\def\ko{k_{\scriptscriptstyle 0}}

\def\eps{\epsilon}

\def\uk{\hat{\#u}_k}

\def\ua{\hat{\#u}_1}
\def\ub{\hat{\#u}_2}
\def\uc{\hat{\#u}_3}

\def\blue{\textcolor{black}}

\begin{document}

\begin{center}

{\bf {\LARGE Controlling Voigt waves by the Pockels effect\footnote{This manuscript is an extension of the conference paper: T.G. Mackay,  ``Voigt waves in electro--optic homogenized composite materials," \emph{SPIE Optics \&
Photonics 2014: Nanostructured Thin Films VII}, San Diego, USA. %; A. Lakhtakia, T.G. Mackay,
% and  M. Suzuki, editors; Proceedings of SPIE {\bf XXXX}, YYYY (2014).
 }
}}

\vspace{10mm} \large

 Tom G. Mackay\footnote{E--mail: T.Mackay@ed.ac.uk}\\
{\em School of Mathematics and
   Maxwell Institute for Mathematical Sciences\\
University of Edinburgh, Edinburgh EH9 3JZ, UK}\\
and\\
 {\em NanoMM~---~Nanoengineered Metamaterials Group\\ Department of Engineering Science and Mechanics\\
Pennsylvania State University, University Park, PA 16802--6812,
USA}\\

\end{center}

\vspace{4mm}

\normalsize

\begin{abstract}

Voigt wave propagation was investigated in a
homogenized composite material (HCM) arising from a porous electro--optic host material infiltrated by a fluid of refractive index $n_a$. The constitutive parameters of the HCM were estimated using the extended Bruggeman homogenization formalism. Numerical studies revealed that the directions which support Voigt wave propagation in the HCM could be substantially controlled by means  of an applied dc electric field.
Furthermore, the extent to which this control could be achieved was found to be sensitive to the
porosity of the host material, the shapes, sizes and orientations of the pores, as well as the refractive index $n_a$.
These findings may be particularly significant for potential technological applications of Voigt waves, such as in optical sensing.

\end{abstract}

\vspace{5mm} \noindent  {\bf Keywords:} Voigt waves, electro--optic materials, Bruggeman
homogenization formalism, Pockels effect

\section{Introduction}

The setting for this numerical study is  electromagnetic plane--wave propagation in anisotropic dielectric materials.
Usually in such materials  two  plane waves with different phase speeds and orthogonal polarizations propagate in a given direction \c{BW}.
However, circumstances can be such that  these two plane waves  coalesce to form a single plane wave, known as a Voigt wave \c{Voigt,Panch,Fedorov,Ranganath}.
A key characteristic of a Voigt wave is that its amplitude is linearly dependent upon propagation distance.
\blue{Indeed, this linear dependence on propagation distance could provide the basis for experimental studies and technological applications that exploit Voigt wave propagation.}
Mathematically, the requirements for Voigt wave propagation are met when an eigenvalue of the corresponding plane--wave propagation matrix  has an algebraic multiplicity which exceeds its geometric multiplicity \c{ML_PiO}.

While certain pleochroic minerals can support Voigt wave propagation \c{Voigt}, greater scope for realizing this singular form of propagation is offered by engineered materials \c{Lakh_helicoidal_bianisotropic_98,Berry}. In particular, homogenized composite
materials (HCMs) may be conceptualized which support Voigt wave propagation whereas the component materials from which they arise do not \c{ML03,ML_WRM}. By judicious design of the structure and selection of the
 constitutive properties of the component materials, the directions in which Voigt waves propagate in a HCM may be controlled.
 This ability to control the directions for Voigt wave propagation may be attractive from the point of view of possible technological applications.
Recently the potential that  Voigt waves offer for optical sensing applications was highlighted \c{M2014_JNP}. The scenario considered involved an HCM arising from a porous biaxial--dielectric host material which was infiltrated by a fluid of refractive index $n_a$. Numerical studies showed that the directions which supported  Voigt wave propagation in the HCM could be acutely sensitive to the refractive index $n_a$; indeed, sensitivities of up to $300^{\circ}$ per RIU were reported.

In the present study, we consider the prospect of controlling Voigt wave propagation in a HCM at the post--fabrication stage,
by means of an applied dc electric field. The HCM here arises from two component materials: a porous electro--optic host  material
which is infiltrated by a fluid of refractive index $n_a$. An extended version \c{M_JNP} of the well--established Bruggeman homogenization formalism \c{Ward,EAB} is employed to estimate the constitutive parameters of the HCM.
\blue{The Bruggeman homogenization formalism was chosen chiefly because (i) there is no restriction on volume fraction of the component materials (unlike in the popular Maxwell Garnett formalism which is restricted to small volume fractions \c{EAB}); and (ii) the shapes and orientations of the particles which make up the two component materials can be independently varied (unlike in the  Maxwell Garnett formalism wherein only the shape and orientation of the designated `inclusion' particles can be varied, and unlike in the strong--permittivity--fluctuation--theory formalism wherein the particles of both component materials are required to have the same shape and orientation \c{EAB}).}

As regards the notation adopted in this paper: vectors are denoted by bold typeface, and the addition of
the $\hat{}$ symbol indicates a unit vector. Accordingly,  the unit vectors aligned with the Cartesian
axes are written as $\hat{\#x}$, $\hat{\#y}$, and $\hat{\#z}$.
 Normal typeface combined with double underlining signifies a  3$\times$3 dyadic. Thus,
   $\=I = \hat{\#x} \, \hat{\#x} + \hat{\#y} \, \hat{\#y} +
\hat{\#z} \, \hat{\#z}$ is the identity 3$\times$3 dyadic and $\=0$ is the null 3$\times$3 dyadic. The dyadic transpose is identified by the superscript $T$. The determinant of a dyadic is delivered by the dyadic operator `$\mbox{det}$'. Blackboard bold typeface combined with double underlining signifies
 a 6$\times$6 dyadic.
 The
permittivity and permeability of free space
are represented by  the symbols $\epso$
and $\muo$, respectively; the free-space wavenumber is expressed as $\ko = \omega
\sqrt{\epso \muo}$ where $\omega$ is the angular frequency.

 \section{Voigt waves}

Our study concerns the propagation of Voigt waves in a homogenized composite material (HCM). As described later in \S\ref{HCM_section}, here the HCM is an anisotropic dielectric material whose electromagnetic properties are characterized by the symmetric dyadic $\=\eps_{\,HCM}$. In general, there are two directions in which Voigt waves may propagate in the HCM. We determine these two directions in an indirect fashion:  Voigt wave propagation is considered along the $z$ coordinate axis for all possible orientations of the HCM. Accordingly, we introduce
  the  HCM
permittivity dyadic in a rotated coordinate frame
\begin{eqnarray}
\={\tilde{\eps}}_{\,HCM} (\alpha, \beta, \gamma ) &=&
 \=R_{\,z}(\gamma)\.\=R_{\,y}(\beta)\.\=R_{\,z}(\alpha)\.\=\eps_{\,HCM}\.
 \=R^T_{\,z}(\alpha)\.\=R^T_{\,y}(\beta)\.\=R^T_{\,z}(\gamma) \l{HCMrot2} \\
 &\equiv& \epso \Big[\tilde{\eps}_{11} \,
\hat{\#x} \, \hat{\#x} + \tilde{\eps}_{22} \, \hat{\#y} \,  \hat{\#y} +
\tilde{\eps}_{33} \, \hat{\#z} \,  \hat{\#z} +  \tilde{\eps}_{12} \, \le \hat{\#x}
\, \hat{\#y} + \hat{\#y} \,  \hat{\#x} \ri \nonumber \\ &&  +
\tilde{\eps}_{13} \, \le \hat{\#x} \, \hat{\#z} + \hat{\#z} \,  \hat{\#x}
\ri + \tilde{\eps}_{23} \, \le \hat{\#y} \, \hat{\#z} + \hat{\#z} \,
\hat{\#y} \ri \Big],
 \l{HCMrot1}
\end{eqnarray}
 where
the orthogonal rotation dyadics
\begin{equation}
\left.
\begin{array}{l}
\=R_{\,y} (\nu)= \cos \nu \le  \, \hat{\#x} \, \hat{\#x} +
\hat{\#z} \, \hat{\#z} \, \ri +
 \sin \nu
\le  \, \hat{\#z} \, \hat{\#x} - \hat{\#x} \, \hat{\#z} \, \ri +
\hat{\#y} \, \hat{\#y} \vspace{4pt} \\
\=R_{\,z} (\nu) =
 \cos \nu
\le  \, \hat{\#x} \, \hat{\#x} + \hat{\#y} \, \hat{\#y} \, \ri +
 \sin \nu
\le  \, \hat{\#x} \, \hat{\#y} - \hat{\#y} \, \hat{\#x} \, \ri +
\hat{\#z} \, \hat{\#z}
\end{array}
\right\}
,
\end{equation}
and  $\alpha$, $\beta$, and $\gamma$ are the three Euler angles
\c{Arfken}.

For the material characterized by the permittivity dyadic $\={\tilde{\eps}}_{\,HCM} (\alpha, \beta, \gamma )$,
Voigt wave propagation parallel to the $z$ axis is supported provided that \c{GL01}:
\begin{itemize}  \item[(i)] $Y (\alpha, \beta, \gamma ) = 0\:$  and
\item[(ii)] $W (\alpha, \beta, \gamma ) \neq 0$,
\end{itemize}
where the scalar quantities
\begin{eqnarray}
Y(\alpha, \beta, \gamma ) &=& \tilde{\eps}^4_{13} + \tilde{\eps}^4_{23} -2
\tilde{\eps}_{23}\tilde{\eps}_{33}
 \les \, 2 \tilde{\eps}_{12}
\tilde{\eps}_{13} - \le \, \tilde{\eps}_{11} - \tilde{\eps}_{22}\, \ri \tilde{\eps}_{23}\,\ris +
\les \le \, \tilde{\eps}_{11}-\tilde{\eps}_{22}\,\ri^2 + 4 \tilde{\eps}^2_{12}\,\ris \,
\tilde{\eps}^2_{33} \nonumber \\ && + 2 \tilde{\eps}_{13} \lec \, \tilde{\eps}^2_{23}
\tilde{\eps}_{13} - \les \, 2 \tilde{\eps}_{12}\tilde{\eps}_{23} + \le \, \tilde{\eps}_{11} -
\tilde{\eps}_{22} \, \ri \, \tilde{\eps}_{13}\,\ris \tilde{\eps}_{33}\,\ric
\end{eqnarray}
and
\begin{equation}
W(\alpha, \beta, \gamma ) = \tilde{\eps}_{12} \tilde{\eps}_{33} - \tilde{\eps}_{13}
\tilde{\eps}_{23}\,.
\end{equation}
These two necessary and sufficient conditions emerge from an eigenanalysis of
the corresponding propagation matrix for plane--wave propagation \c{GL01}.
 In particular, these conditions can be satisfied by certain biaxial dielectric materials but not by
isotropic or  uniaxial  dielectric materials.

 \section{Extended Bruggeman homogenization formalism}

 \subsection{Component materials}

 Our study is based on the homogenization of two particulate component materials, namely component material $a$ and component material $b$. Component material $a$ is simply envisaged as an isotropic dielectric fluid characterized by the permittivity dyadic $\=\eps^{(a)} = \epso \eps^{(a)} \=I$ in the optical regime; its relativity permittivity $\eps^{(a)} = n_a^2$. Component material $b$ is an electro-optic material that exhibits the Pockels effect in the optical regime.
\blue{ Its linear electro-optic properties
are conventionally characterized via its inverse   permittivity dyadic \c{Boyd,LM07}.
Retaining only the first--order components of the dc electric field (which is the usual approximation in electro--optics  \c{Yariv_Yeh}), the permittivity dyadic for component material $b$ may be written as
\c{L06_JEOS}}
\begin{eqnarray}
\nonumber \=\eps^{(b)} &\approx& \epso
\Bigg\{
\sum_{k=1}^3\les\epsilon _{k}^{(b)}\le
1-\epsilon _{k}^{(b)}s_k \ri\,\uk\uk\ris\\[5pt]
&-&\epsilon _{2}^{(b)}\epsilon _{3}^{(b)}\,s_4 \le\ub\uc
+\uc\ub\ri -\epsilon _{1}^{(b)}\epsilon _{3}^{(b)}\,s_5
\le\ua\uc +\uc\ua\ri -\epsilon _{1}^{(b)}\epsilon
_{2}^{(b)}\,s_6 \le\ua\ub +\ub\ua\ri \Bigg\}, \label{PocEps}
\end{eqnarray}
 under the assumption that
\begin{equation}
\label{restriction}
\frac{1}{\epso}
\lec \lvert \eps_1^{(b)} \rvert, \lvert \eps_2^{(b)} \rvert, \lvert \eps_3^{(b)} \rvert \ric_{\mbox{max}}\,\,
 \lec \lvert s_1 \rvert, \lvert s_2 \rvert, \lvert s_3 \rvert, \lvert s_4 \rvert, \lvert s_5 \rvert, \lvert s_6 \rvert
 \ric_{\mbox{max}}\ll 1\,.
\end{equation}
In Eq.~\r{PocEps} the orientations of the unit vectors
\begin{equation}
\left.\begin{array}{l} \ua=-(\hat{\#x}\cos\phi_b+\hat{\#y}\sin\phi_b)
\cos\theta_b+\hat{\#z}\sin\theta_b\\[5pt]
\ub=\hat{\#x}\sin\phi_b-\hat{\#y}\cos\phi_b\\[5pt]
\uc=(\hat{\#x}\cos\phi_b+\hat{\#y}\sin\phi_b)\sin\theta_b+\hat{\#z}\cos\theta_b
\end{array}\right\}\,,
\end{equation}
as specified by the angles $\theta_b\in\les0,\pi\ris$ and $\phi_b\in\les0,2\pi\ris$,
are determined by the crystallographic structure of the material. The dependency of component material $b$'s constitutive parameters  upon a uniform dc electric field $\#E^{dc}$ is captured by the scalar parameters
\begin{equation}
\label{sJ} s_j= \sum_{k=1}^3 r_{jk}\,\uk\.\#E^{dc}\,,\quad \qquad
j\in\lec 1,2,3,4,5,6 \ric,
\end{equation}
which are
expressed in terms of the electro-optic coefficients  $r_{jk}$ ($j\in\lec1,2,3,4,5,6\ric$,
$k\in\le1,2,3\ric$).
Thus, depending upon the relative values of the principal permittivity scalars $\epsilon _{1,2,3}^{(b)}$, component material $b$ may be an isotropic, uniaxial or biaxial material; furthermore, depending upon the relative values of the electro--optic coefficients, it may belong to one of 20 crystallographic classes of point
group symmetry \c{EAB}.

Both component materials are composed of spheroidal particles. These component spheroids are randomly distributed
 with the spheroids of component material $a$ occupying the volume fraction $f_a$ and those of
 component material $b$ occupying the volume fraction $f_b = 1-f_a$.
 It is assumed  that all component $a$ spheroids have the same shape and orientation, and  all component $b$ spheroids have the same shape and orientation. The surface of each component spheroid, relative to its centroid, is prescribed by the position vector
\begin{equation} \l{er}
\#r_\ell = \eta_\ell \, \=U_{\,\ell} \. \hat{\#r}, \qquad (\ell = a,
b).
\end{equation}
Here the position vector $\hat{\#r}$ prescribes the surface of the unit sphere; the shape and orientation of the component spheroid
are encapsulated by the real--symmetric dyadic $\=U_{\,\ell}$; and the linear dimensions of the component spheroid
are characterized by the size parameter $\eta_\ell > 0$.
\blue{ The homogenization of component materials $a$ and $b$ is founded on the notion  that  $\eta_\ell$ is much smaller than the
wavelengths involved. But, in extended homogenization formalisms, $\eta_\ell$ need not be vanishingly small \cite{M_JNP}.
Accordingly, in the following numerical studies, the range $0.2 <
\ko \eta_\ell
< 0.3 $ is used.}
For simplicity, we assume that
 $\eta_a \equiv \eta_b$; and henceforth  $\eta$ is written in lieu of $\eta_\ell$ $ (\ell = a,
b)$.

For definiteness, we
 choose the axis of rotational symmetry for the component $b$ spheroids to be aligned with the $x$ coordinate axis.
Thus, the surface dyadic for
the component material $b$ spheroids may be expressed as
\begin{equation}
\=U_{\,b} = \frac{1}{\sqrt[3]{U_x U^2 }} \les \, U_x \hat{\#x} \,
\hat{\#x} + U \le  \hat{\#y} \, \hat{\#y} +  \hat{\#z} \, \hat{\#z}
 \ri \ris  , \qquad (U_x, U > 0).
\end{equation}
The axis of rotational symmetry for the component material $a$ spheroids is rotated
by an angle $\varphi$  in the $xy$ plane relative  to the
axis of rotational symmetry for the component material $b$ spheroids.
Thus, the surface dyadic for
the component material $a$ spheroids may be expressed as
 \begin{equation}
\=U_{\,a} =  \=R_{\,z} (\varphi) \. \=U_{\,b}   \. \=R^T_{\,z} (\varphi).
\end{equation}

 \subsection{Homogenized composite material} \l{HCM_section}

In the long--wavelength regime, the random mixture of component materials $a$ and $b$ may be regarded as a homogenized composite material (HCM). The electromagnetic properties of this HCM  are characterized by its symmetric permittivity dyadic $\=\eps_{HCM}$. We utilize the extended Bruggeman formalism \c{M_JNP,Goncharenko} to estimate $\=\eps_{HCM}$.
This approach is based on the nonlinear dyadic equation
\c{WLM97}
\begin{equation} \l{Bruggeman_eqn}
f_a \lec \le \=\eps^{(a)} -  \=\eps_{\,HCM} \ri \. \les \=I +
\=D_{\,a} \. \le \=\eps^{(a)} -  \=\eps_{\,HCM} \ri \ris^{-1} \ric +
f_b \lec \le \=\eps^{(b)} -  \=\eps_{\,HCM} \ri \. \les \=I +
\=D_{\,b} \. \le \=\eps^{(b)} -  \=\eps_{\,HCM} \ri \ris^{-1} \ric =
\=0\,.
 \end{equation}
 The depolarization dyadics $\=D_{\,a,b}$ herein encapsulate the electromagnetic responses of $\=U_{\,a,b}$--shaped spheroids
 embedded in the HCM. Further details are provided in the Appendix.

\section{Numerical investigations}

The question that we now turn to is: to what extent can the directions that support Voigt wave propagation
be influenced by the application of a  dc electric field $\#E_{dc}$? We do so by means of representative numerical calculations.

\subsection{Component materials}

We consider a porous electro--optic host material $b$, characterized by the permittivity dyadic $\=\eps^{(b)}$, which is infiltrated by a fluid of refractive index $n_a$. For component material $b$, we choose potassium niobate which is specified by
 \c{ZSB}:
 $\epsilon_1^{(b)} = 4.72$, $\epsilon_2^{(b)}= 5.20$,
$\epsilon_3^{(b)}=5.43$, $r_{13}=34\times 10^{-12}$~m~V$^{-1}$,
$r_{23}=6\times 10^{-12}$~m~V$^{-1}$, $r_{33}=63.4\times
10^{-12}$~m~V$^{-1}$, $r_{42}=450\times 10^{-12}$~m~V$^{-1}$,
$r_{51}=120\times 10^{-12}$~m~V$^{-1}$, and all other
$r_{jk}\equiv0$. And for component material $a$ we consider the range $n_a \in \les 1.05, 1.3 \ris$. The eccentricity parameter $\rho$ is introduced as a gauge of the shape of the component spheroids: we let $U_x = 1 + \rho$, $U = 1 - \le\rho/20\ri$, and consider the range $\rho \in \les 2,4 \ris$.

In order to emphasize the degree of electrical control that may be attained over the Voigt wave directions, we explore cases where the Pockels effect is most easily discernable. To this end, the principal crystallographic axis of component material $b$ is chosen to be aligned with its component spheroids; in addition, the  dc electric field $\#E_{dc}$ is also chosen to be aligned with the component $b$ spheroids.  Accordingly, the crystallographic angles $\theta_b = \phi_b =
0$; and we have  $\#E^{dc} = E^{dc}_{3} \hat{\#u}_3$ with \blue{the range for the electric field component  $E^{dc}_3 $ being such that the inequality
\r{restriction} is satisfied}.

\subsection{Extended Bruggeman estimates of $\=\epsilon_{\,HCM}$}

\begin{figure}[!h]
\centering \psfull \epsfig{file=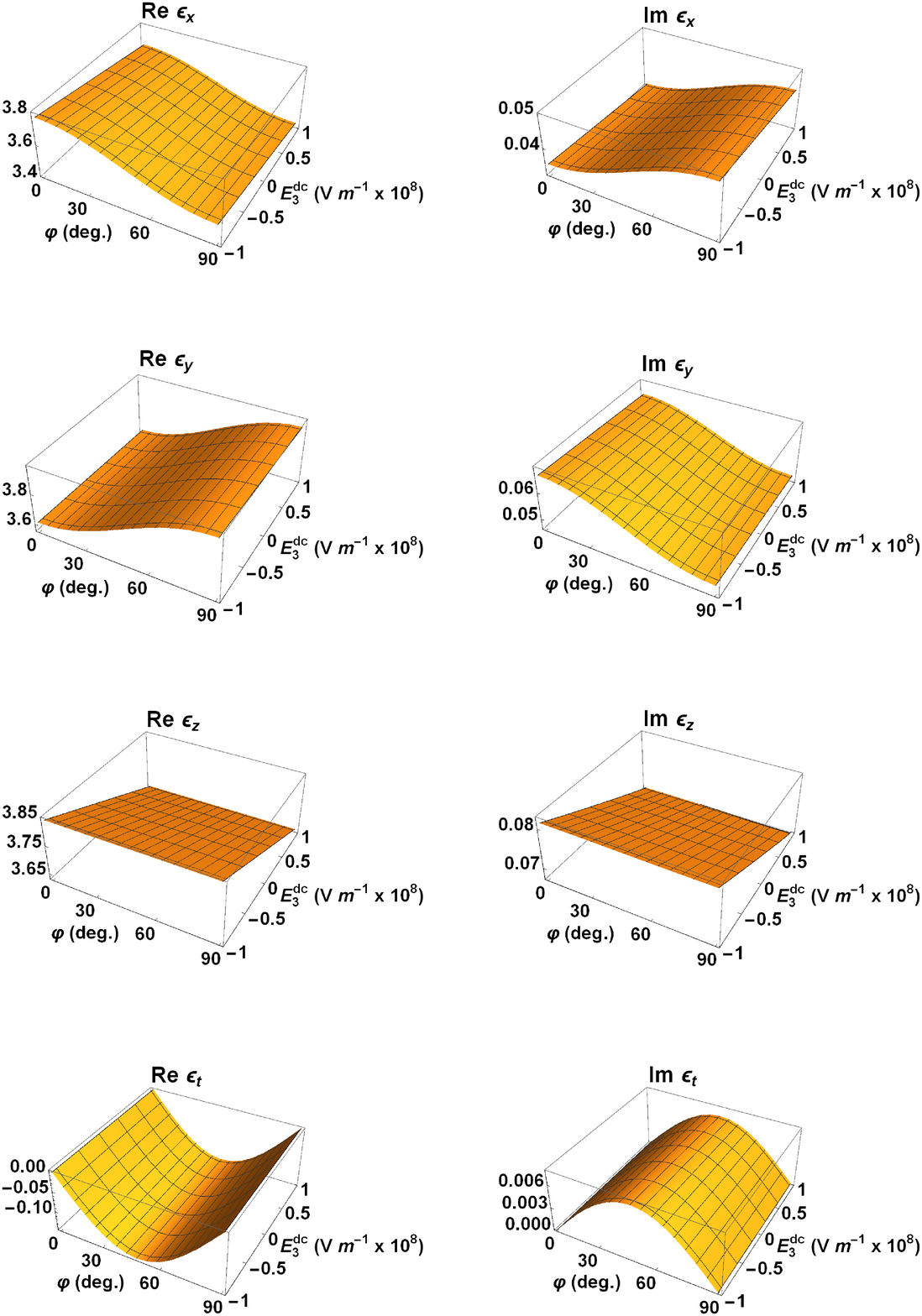,width=5.0in}
 \caption{The extended Bruggeman estimates of relative permittivity parameters of the HCM plotted versus
 \blue{ $E^{dc}_3 \in \le -1, 1 \ri \times 10^8 $ $\mbox{V} \mbox{m}^{-1} $}
 and spheroid orientation angle $\varphi \in \le 0, 90 \ri^\circ$.
The refractive index  $n_a = 1.1$, volume fraction $f_a = 0.3$,  size parameter $ \eta = 0.3/\ko$, and eccentricity
parameter $\rho = 3$. } \label{fig1}
\end{figure}

Before exploring the directions which support Voigt wave propagation in relation to a  dc electric field,
we first consider how such an electric field effects the constitutive parameters of the HCM.
For the constitutive parameter values chosen for our numerical investigations, the HCM is a biaxial dielectric material characterized by  a symmetric permittivity
dyadic of the general form \cite{MW_biax2}
\begin{equation} \l{e_HCM}
\=\eps_{\,HCM} = \epso \les \, \eps_x   \, \hat{\#x} \, \hat{\#x} +
\eps_y \hat{\#y} \, \hat{\#y}  +
 \eps_t \le  \hat{\#x} \, \hat{\#y} + \hat{\#y} \, \hat{\#x} \, \ri + \eps_z \hat{\#z}
\, \hat{\#z} \, \ris.
\end{equation}
In Fig.~\ref{fig1}
the real and imaginary parts of the  extended Bruggeman estimates of the relative permittivity parameters $\eps_{x,y,z,t}$ are plotted versus
the dc electric field component $E^{dc}_{3}$, and the spheroid orientation angle $\varphi$. For these calculations, the volume faction $f_a = 0.3$, the refractive index $n_a = 1.1$, the size parameter $\eta = 0.3/\ko$, and the eccentricity
parameter $\rho = 3$. We see that both the real and imaginary parts of  $\eps_{x,y,z}$ decrease in an approximately linear manner as $E^{dc}_{3}$ increases from \blue{ $-1 \times 10^8$  $\mbox{V} \mbox{m}^{-1} $ to $1 \times 10^8$  $\mbox{V} \mbox{m}^{-1} $, albeit this effect is much more obvious  for $\eps_z$ than it is for $\eps_{x,y}$}. The effect of $E^{dc}_{3}$ on both the real and imaginary parts of  $\eps_{t}$   is barely discernable in Fig.~\ref{fig1}.
Both the real and imaginary parts of $\eps_{x,y}$ vary markedly as $\varphi$ increases from $0^\circ$ to $90^\circ$.
However, the relative permittivity parameter $\eps_z$ seems to be largely unaffected by the rotation in the $xy$ plane represented by $\varphi$.
The off-diagonal relative permittivity parameter $\eps_t = 0$ at $\varphi = 0^\circ$ and $90^\circ$;  both the real and imaginary parts of $\eps_t $ increase and then decrease markedly as
 $\varphi$ increases from  $0^\circ$ to $90^\circ$.

In addition to the dc electric field and the orientation of the component spheroids,
the electromagnetic properties of HCM are also sensitive to the shapes and sizes of the component spheroids, the volume fraction, and the constitutive parameters of the component materials. These matters have been explored using the extended Bruggeman homogenization formalism in earlier studies, to which the reader is referred for further details
\c{ML_JAP_2007,M2011_JOPA}.

\subsection{Electrical control over directions that support Voigt wave propagation}

Now we investigate the effects of a dc electric field on the  directions which support Voigt wave propagation in the HCM.
These Voigt wave directions are represented by the
Euler angles
$\alpha$, $\beta$, and $\gamma$ for which $Y (\alpha, \beta, \gamma ) = 0$ and $W (\alpha, \beta, \gamma ) \neq 0$.
Notice that
 the angular coordinate $\gamma$ may be eliminated from our investigations since
 Voigt wave propagation parallel to the $z$ axis (in the
rotated coordinate system) is independent of rotation about that
axis. For the HCM specified by a permittivity dyadic of the form given in Eq.~\r{e_HCM}, there are generally two distinct orientations for which
 the Voigt wave conditions $Y=0$ and $W \neq 0$ are
satisfied. We label these two orientations with the angular
coordinates $\alpha = \alpha_{1,2}$ and $\beta = \beta_{1,2}$; the corresponding values of the quantity $W$ are written as $W_{1,2}$.

\subsubsection{Orientation of component spheroids} \l{Sec_orient}

\begin{figure}[!h]
\centering \psfull \epsfig{file=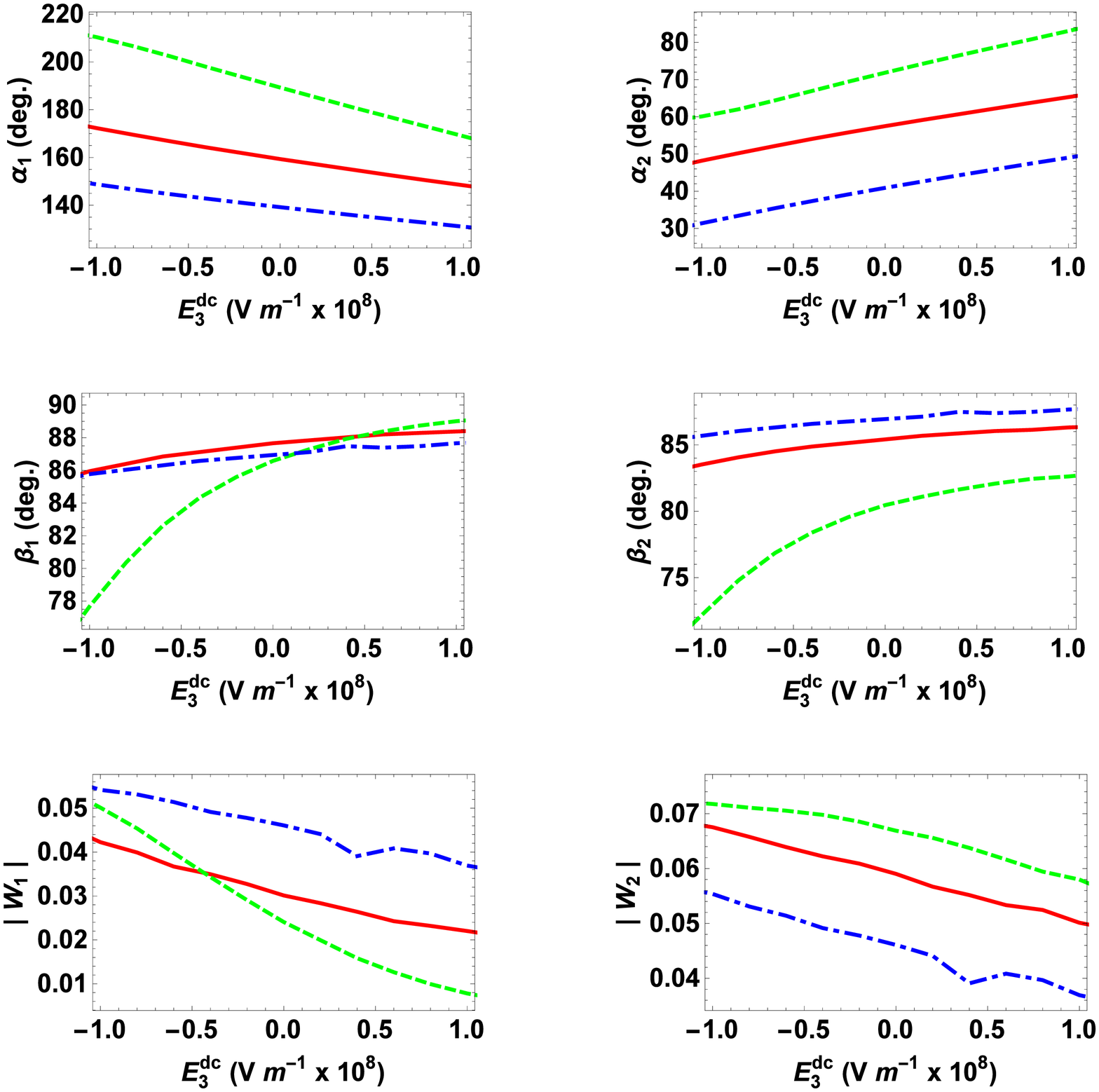,width=5.8in}
 \caption{The angles $\alpha_{1,2}$ and $\beta_{1,2}$, along with the quantities $| W_{1,2} |$, plotted versus
 \blue{  $E^{dc}_3 \in \le -1, 1 \ri \times 10^8 $ $\mbox{V} \mbox{m}^{-1} $} for the spheroid orientation angles $\varphi = 60^\circ$ (red, solid curves), $30^\circ$ (green, dashed curves), and $90^\circ$ (blue, broken dashed curves).
The refractive index  $n_a = 1.1$, volume fraction $f_a = 0.3$,   size parameter $\eta = 0.3/\ko$, and eccentricity
parameter $\rho = 3$. } \label{fig2}
\end{figure}

We begin by considering the effect of the  dc electric field on the Voigt wave directions in relation to the orientation of the component spheroids. In Fig.~\ref{fig2}, the angles $\alpha_{1,2}$ and $\beta_{1,2}$ along with the quantities $| W_{1,2} |$ are plotted versus $E^{dc}_{3}$ for three different orientations of the component material $a$ spheroids as given by
 $\varphi = 30^\circ$ (green, dashed curves), $60^\circ$ (red, solid curves),  and $90^\circ$ (blue, broken dashed curves).
Here the refractive index  $n_a = 1.1$, the volume fraction $f_a = 0.3$,  the size parameter $ \eta = 0.3/\ko$ and the eccentricity
parameter $\rho = 3$.
Clearly the directions that support Voigt wave propagation are highly sensitive to the  dc electric field, for all orientations of the component material $a$ spheroids considered. \blue{Specifically, the $\alpha$ angular coordinate varies by as much as $38^\circ$, and the $\beta $ angular coordinate varies by as much as $12^\circ$, as the value of
 $E^{dc}_{3}$ ranges from $-1 \times 10^8$  $\mbox{V} \mbox{m}^{-1} $ to $1 \times 10^8$  $\mbox{V} \mbox{m}^{-1} $.}
Generally,  greater changes in the Voigt wave directions are observed for smaller values of the spheroid orientation angle $\varphi$.  \blue{Also, the  angular coordinate $\beta$ varies relatively little for
$E^{dc}_{3} \in \le 0, 1 \ri  \times 10^8$  $\mbox{V} \mbox{m}^{-1} $, but varies  considerably more for
$E^{dc}_{3} \in \le -1, 0 \ri  \times 10^8$  $\mbox{V} \mbox{m}^{-1} $.}
We note that the quantities $| W_{1,2} |$ are nonzero for all calculations, in accordance with requirements for Voigt wave propagation. Other than being nonzero, the quantities $| W_{1,2} |$ are of no obvious physical significance; therefore, plots of these quantities shall be omitted henceforth.

\subsubsection{Shape of component spheroids}

We repeat the calculations of \S\ref{Sec_orient} but now fixing the orientation of the component material $a$ spheroids
at $\varphi = 60^\circ$ and allowing the eccentricity parameter $\rho$ to vary. The results are presented in Fig.~\ref{fig3}
where the angles $\alpha_{1,2}$ and $\beta_{1,2}$  are plotted versus $E^{dc}_{3}$ for
  $\rho = 2$ (green, dashed curves),   $ 3$ (red, solid curves), and $4$ (blue, broken dashed curves).
  The sensitivity of the Voigt wave directions to the dc electric field is significantly modulated by the degree of eccentricity of the component spheroids. \blue{Specifically, the angular coordinates $\alpha_{1,2}$ are moderately sensitive to $\rho$. While the angular coordinates $\beta_{1,2}$ are relatively insensitive to
   $\rho$ for $E^{dc}_{3} \in \le 0, 1 \ri  \times 10^8$  $\mbox{V} \mbox{m}^{-1} $, these angles are highly sensitive to $\rho$ for  $E^{dc}_{3} \in \le -1, 0 \ri  \times 10^8$  $\mbox{V} \mbox{m}^{-1} $.} The influence of the shape parameter upon the sensitivity of the Voigt wave directions in relation to $E^{dc}_{3}$ appears to be greatest for the smallest value of $\rho$
considered here.

\subsubsection{Size of component spheroids}

\begin{figure}[!h]
\centering \psfull \epsfig{file=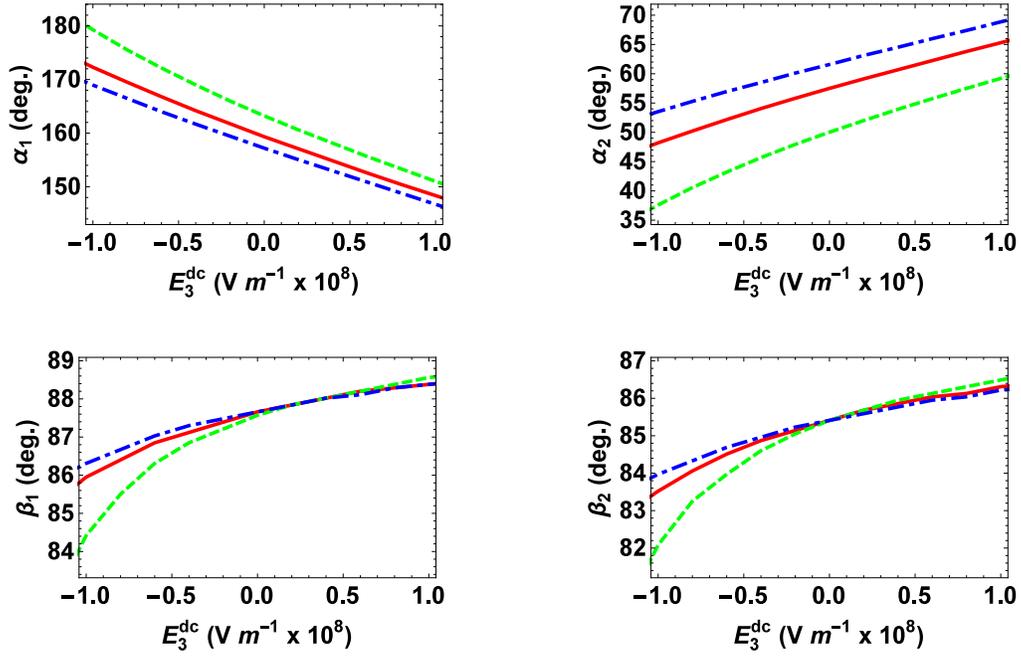,width=5.8in}
 \caption{The angles $\alpha_{1,2}$ and $\beta_{1,2}$ plotted versus
  \blue{ $E^{dc}_3 \in \le -1, 1 \ri \times 10^8 $ $\mbox{V} \mbox{m}^{-1}$ } for the eccentricity parameter values
    $\rho = 3$ (red, solid curves), $2$ (green, dashed curves), and $4$ (blue, broken dashed curves).
The refractive index  $n_a = 1.1$, volume fraction $f_a = 0.3$,  size parameter $\eta = 0.3/\ko$, and spheroid orientation angle $\varphi = 60^\circ$. } \label{fig3}
\end{figure}

\begin{figure}[!h]
\centering \psfull \epsfig{file=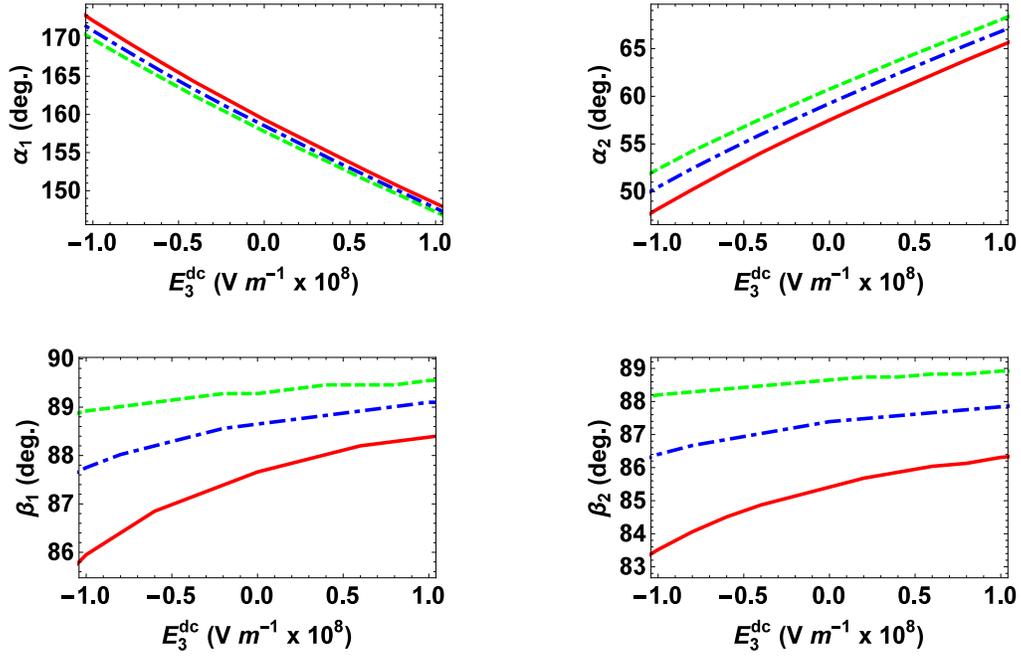,width=5.8in}
 \caption{The angles $\alpha_{1,2}$ and $\beta_{1,2}$  plotted versus
\blue{   $E^{dc}_3 \in \le -1, 1 \ri \times 10^8 $ $\mbox{V} \mbox{m}^{-1} $ } for the
    size parameter values $ \eta = 0.3/\ko$
     (red, solid curves), $0.2/\ko$ (green, dashed curves), and $0.25/\ko$ (blue, broken dashed curves).
The refractive index  $n_a = 1.1$,  volume fraction $f_a = 0.3$,  spheroid orientation angle $\varphi = 60^\circ$, and eccentricity
parameter $\rho = 3$. } \label{fig4}
\end{figure}

Next we turn to the size of the component spheroids, as gauged by the size parameter $\eta$. In Fig.~\ref{fig4} plots similar to those in Fig.~\ref{fig3} are provided but here the eccentricity parameter is fixed at $\rho = 3$ and three different values of size parameter are employed, namely $\eta  =
0.2/\ko$ (green, dashed curves),
 $0.25/\ko$ (blue, broken dashed curves), and
$0.3/\ko$
     (red, solid curves).
The general trends observable in Fig.~\ref{fig4} are similar to those observable in Fig.~\ref{fig3}. That is,
the size parameter does significantly modulate the dependency of the Voigt wave directions on the dc electric field; the influence of the size parameter is relatively modest for the $\alpha_{1,2}$ angular coordinates but is considerably stronger for the $\beta_{1,2}$ angular coordinates, especially at lower values of $E^{dc}_{3}$.

\subsubsection{Volume fraction}

\begin{figure}[!h]
\centering \psfull \epsfig{file=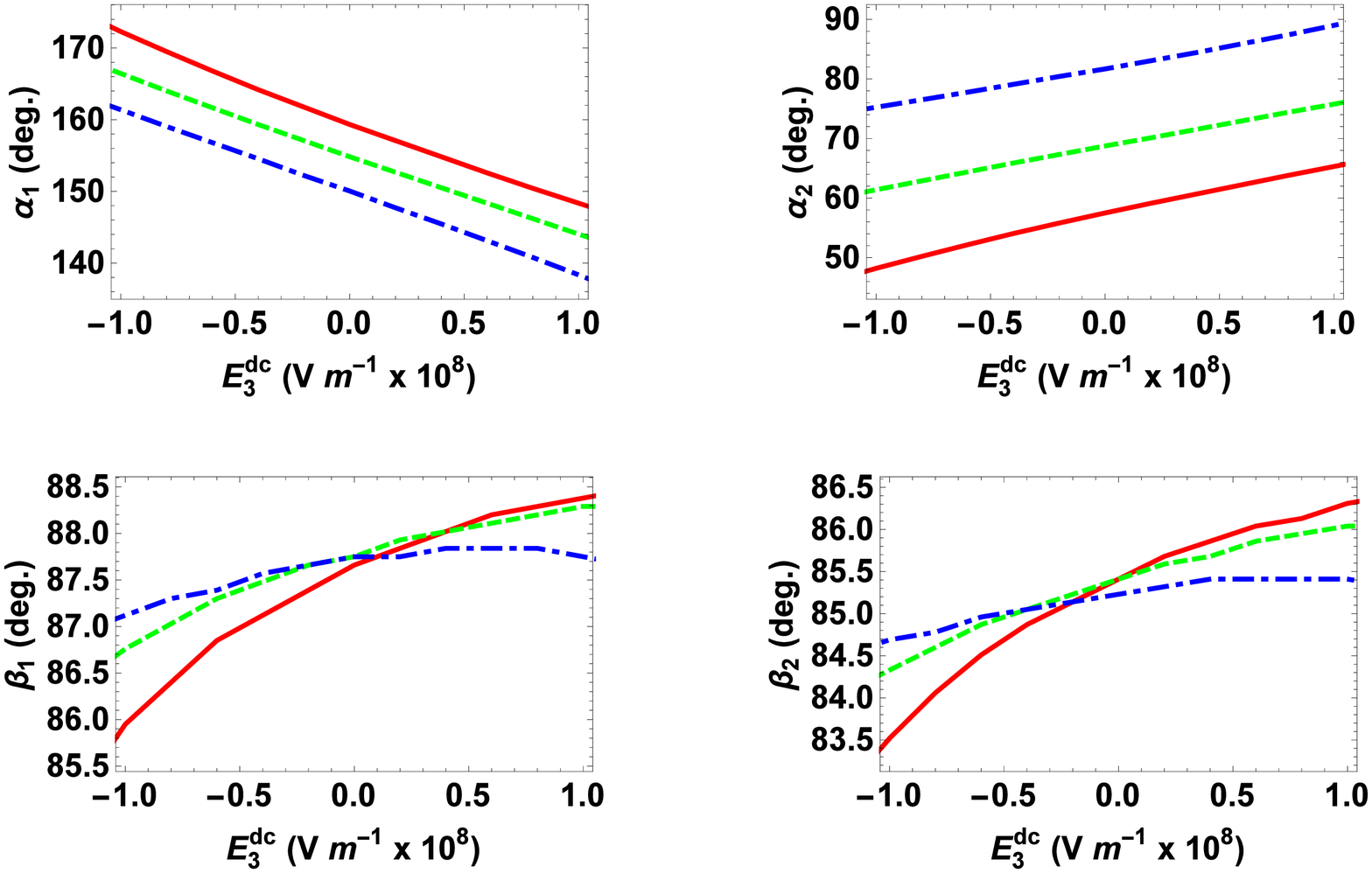,width=5.8in}
 \caption{The angles $\alpha_{1,2}$ and $\beta_{1,2}$ plotted versus
\blue{   $E^{dc}_3 \in \le -1, 1 \ri \times 10^8 $ $\mbox{V} \mbox{m}^{-1}$ } for the volume fractions $f_a = 0.3$ (red, solid curves), $0.35$ (green, dashed curves), and $0.4$ (blue, broken dashed curves).
The refractive index  $n_a = 1.1$,  size parameter $\eta = 0.3/\ko$,  spheroid orientation angle $\varphi = 60^\circ$, and eccentricity
parameter $\rho = 3$. } \label{fig5}
\end{figure}

The influence of the volume fraction occupied by component material $a$ (or, equivalently, the porosity of component material $b$) is now considered. Plots similar to those in Fig.~\ref{fig4} are provided in Fig.~\ref{fig5} but here
the size parameter $ \eta = 0.3/\ko$; results are plotted for the volume fractions
 $f_a = 0.3$ (red, solid curves), $0.35$ (green, dashed curves), and $0.4$ (blue, broken dashed curves).
The approximate linearity of the relationship between the $\alpha_{1,2}$ angular coordinates and $E^{dc}_{3}$ is
not greatly affected by the volume fraction. Furthermore, in the vicinity of
$E^{dc}_{3} \approx 0  \times 10^8$ $\mbox{V} \mbox{m}^{-1} $, the angular coordinates $\beta_{1,2}$ seem to be insensitive to the volume fraction. But for larger and smaller values of $E^{dc}_{3}$, the volume faction has a strong influence on the angular coordinates $\beta_{1,2}$ in relation to  $E^{dc}_{3}$. This influence is strongest at smaller values of $f_a$ when  $E^{dc}_{3}$ is negative--valued and at larger values of $f_a$ when  $E^{dc}_{3}$ is positive--valued.

\subsubsection{Refractive index of infiltrating fluid}

\begin{figure}[!h]
\centering \psfull \epsfig{file=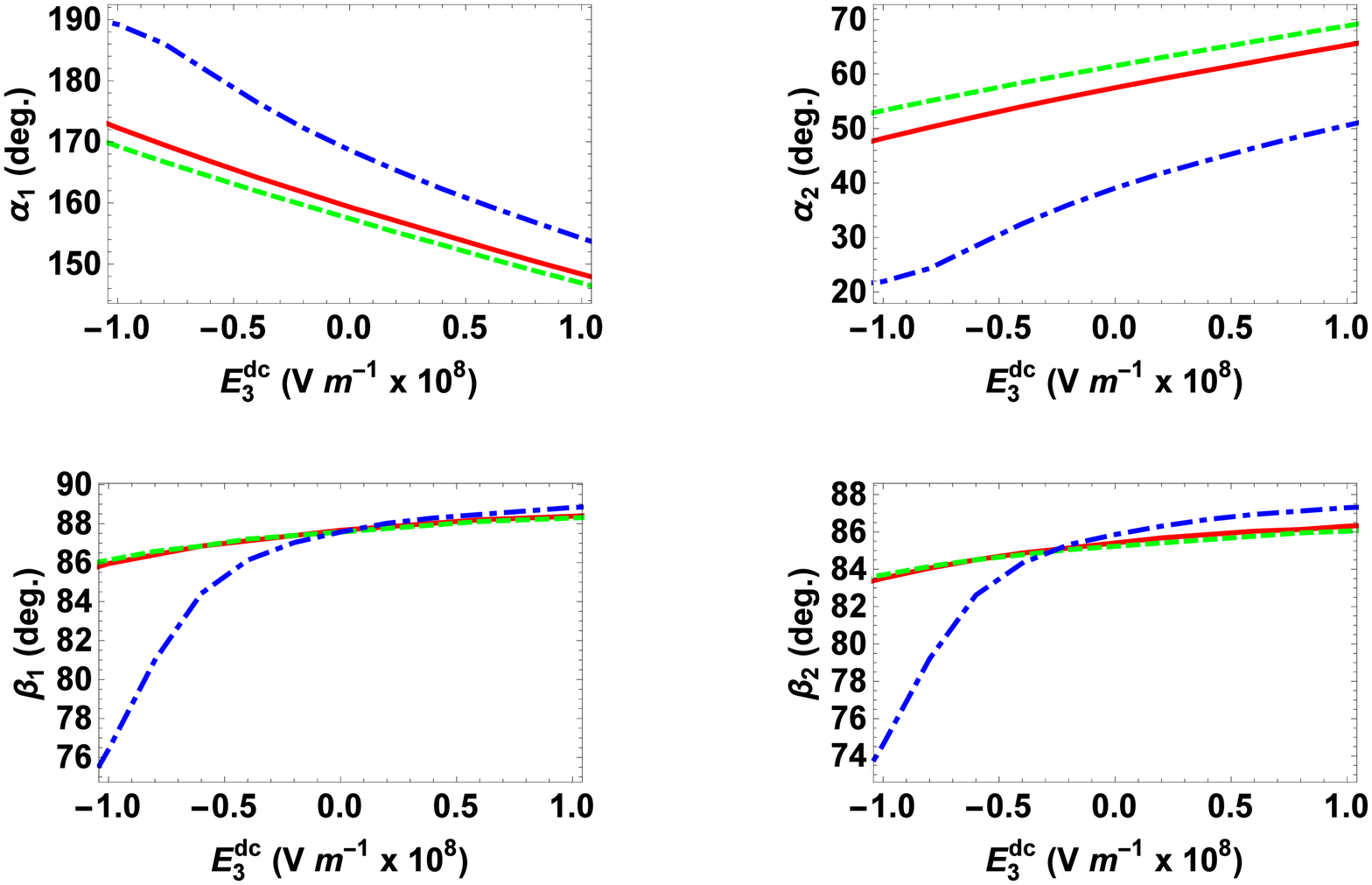,width=5.8in}
 \caption{The angles $\alpha_{1,2}$ and $\beta_{1,2}$ plotted versus
 \blue{  $E^{dc}_3 \in \le -1, 1 \ri  \times 10^8 $ $\mbox{V} \mbox{m}^{-1} $} for the refractive index values
    $n_a = 1.1$ (red, solid curves), $1.05$ (green, dashed curves), and $1.3$ (blue, broken dashed curves).
The  volume fraction $f_a = 0.3$,  size parameter $\eta = 0.3/\ko$,  eccentricity
parameter $\rho = 3$,  and  spheroid orientation angle $\varphi = 60^\circ$. } \label{fig6}
\end{figure}

Lastly, let us consider the influence of the refractive index of component material $a$, namely $n_a$.
We repeat the calculations of Fig.~\ref{fig5} but with the volume fraction fixed at $f_a = 0.3$; results are plotted in Fig.~\ref{fig6} for the refractive index values  $n_a = 1.05$ (green, dashed curves),
    $ 1.1$ (red, solid curves), and $1.3$ (blue, broken dashed curves).
It may be seen that the refractive index strongly modulates the sensitivity of the  Voigt wave directions to the dc electric field. Reminiscent of the general trends in Fig.~\ref{fig3}, \blue{ the refractive index $n_a$ has only a modest influence upon
 the angular coordinates $\beta_{1,2}$
 for $E^{dc}_{3} \in \le 0, 1 \ri  \times 10^8$  $\mbox{V} \mbox{m}^{-1} $, but these angles are highly sensitive to $n_a$ for  $E^{dc}_{3} \in \le -1, 0 \ri  \times 10^8$  $\mbox{V} \mbox{m}^{-1} $.} The effects of the refractive index upon the sensitivity of the Voigt wave directions in relation to $E^{dc}_{3}$ are generally greatest for the smallest value of $n_a$
considered here.

\section{Closing comments}

Voigt wave propagation has been considered for a HCM arising from a porous electro--optic host material, namely potassium niobate, infiltrated by a fluid of refractive index $n_a$. By means of representative numerical calculations, it has been demonstrated that the directions  which support the propagation of Voigt waves  may be substantially controlled by the application of a dc electric field. Furthermore, the extent to which this control may be achieved is dependent upon the
porosity of the host material, the shapes, sizes and orientations of the pores, as well as the refractive index $n_a$.

\blue{
 In order to be physically realistic, the magnitude of the applied dc electric field should be less than the  magnitudes of the breakdown electric fields of the materials involved. However, the breakdown
electric fields depend upon the morphologies and thicknesses of the materials, with thin films generally capable of withstanding higher electric fields than bulk materials. Typical  breakdown fields for potassium niobate thin films are estimated to be about
$8 \times 10^7 \; \mbox{V m}^{-1}$ \c{Madou}, while Kingon \emph{et al.} reported no breakdown when a field of
$1 \times 10^8 \; \mbox{V m}^{-1}$ was applied to a potassium niobate thin film in their experiments \c{Kingon}.
Accordingly, it seems reasonable here to present results for applied  dc electric fields of magnitudes up to $1.0 \times 10^8 \; \mbox{V m}^{-1}$. Parenthetically, the  inequality
\r{restriction} is satisfied for such electric fields.
In a practical device,
the maximum dc electric field that could be applied may well be determined by factors other than the breakdown of the electro--optic material, such as how well the electronic components of the device can be shielded from  the dc electric field. Furthermore, potassium niobate was chosen here as a representative example of an electro--optic material which has relatively large electro--optic coefficients. If artificial electro--optic materials were to be developed  with larger electro--optic coefficients then this would allow the possibility of using dc electric fields of lower magnitude. }

The findings reported here may be particulary significant for potential technological applications of Voigt waves. Notably, Voigt waves appear to present promising opportunities for optical sensing applications, since the directions which support Voigt wave propagation in a porous host material can be highly sensitive to changes in the refractive index of a fluid that infiltrates the host material \c{M2014_JNP}. Thus, it may be envisaged that electrical control could be harnessed to track Voigt wave propagation in an optical--sensor setting.
Other possible applications of electrical control of Voigt waves may involve optical switching \c{Optical_switch}.
Such Voigt wave applications  are matters for future study.

\section*{\blue{Appendix}}

\blue{
 In the extended Bruggeman formalism encapsulated by Eq.~\r{Bruggeman_eqn}, the depolarization dyadics
comprise the sums \c{M_WRM}
\begin{equation}
\=D_{\,\ell} = \=D^{0}_{\,\ell} +  \=D^{+}_{\,\ell}, \qquad (\ell =
a, b).
\end{equation}
The dyadic term $\=D^{0}_{\,\ell}$ represents the depolarization
contribution arising in the limit $\eta \to 0$; it is given by the double integral \c{M97,MW97}
\begin{equation} \l{D_o}
\=D^{0}_{\,\ell} = \frac{1}{4 \pi}\, \int^{2 \pi}_{\phi = 0}
\int^{\pi}_{\theta = 0} \frac{ \le \=U^{-1}_{\, \ell} \. \hat{\#q}
\ri \,  \le \=U^{-1}_{\, \ell} \. \hat{\#q} \ri \; \sin \theta}{\le
\=U^{-1}_{\, \ell} \. \hat{\#q} \ri \. \=\eps_{\, HCM} \. \le
\=U^{-1}_{\, \ell} \. \hat{\#q} \ri}\, d \theta\, d \phi \, , \qquad
(\ell = a, b),
\end{equation}
with the unit vector $\hat{\#q} = \sin \theta \cos \phi \,
\hat{\#x} + \sin \theta \sin \phi \, \hat{\#y} + \cos \theta \,
\hat{\#z}$.
The depolarization contribution associated with the
nonzero size of the component particles is represented by the
 dyadic term $\=D^{+}_{\,\ell}$, which  is conveniently given as a 3$\times$3 subdyadic of the  6$\times$6 dyadic $\underline{\underline{\mathbb{D}}}^{+}_{\,\ell}$ per
\begin{equation}
\les \, \=D^{+}_{\,\ell} \, \ris_{m n} = \les \,
\underline{\underline{\mathbb{D}}}^{+}_{\,\ell} \, \ris_{m n},
\qquad \le m, n \in \lec 1, 2, 3 \ric \ri.
\end{equation}
Here the 6$\times$6 dyadic  \c{M_WRM}
\begin{eqnarray} \l{D_plus}
\underline{\underline{\mathbb{D}}}^{+}_{\,\ell} &=&
\frac{\omega^4}{4 \pi \muo } \int^{2 \pi}_{\phi = 0}
\int^{\pi}_{\theta = 0} \frac{\sin \theta}{\les \le \=U^{-1}_{\,
\ell} \. \hat{\#q} \ri \. \=\eps_{\, HCM} \. \le \=U^{-1}_{\, \ell}
\. \hat{\#q} \ri \ris \le \=U^{-1}_{\, \ell} \. \hat{\#q} \ri \. \le
\=U^{-1}_{\, \ell} \. \hat{\#q} \ri } \nonumber \\ && \times \Bigg[
\frac{1}{ \kappa_+ - \kappa_-  }  \Bigg( \frac{\exp \le i \eta q \ri
}{2 q^2} \le 1 - i \eta q\ri
 \Big\{ \,
 \mbox{det} \les \underline{\underline{\mathbb{A}}} (\=U^{-1}_{\, \ell} \.\#q ) \ris \, \underline{\underline{\mathbb{G}}}^{+} (\=U^{-1}_{\,
\ell} \.\#q) \nonumber \\
&&
  +  \mbox{det} \les \underline{\underline{\mathbb{A}}} (-\=U^{-1}_{\, \ell} \.\#q ) \ris  \,  \underline{\underline{\mathbb{G}}}^{+} (-\=U^{-1}\.\#q
) \Big\} \Bigg)^{q=\sqrt{ \kappa_+ }}_{q=\sqrt{ \kappa_- }} + \frac{
 \mbox{det} \les \underline{\underline{\mathbb{A}}} (\#0 ) \ris} {\kappa_+  \, \kappa_- }\,
\underline{\underline{\mathbb{G}}}^{+} (\#0)
 \Bigg] \; d \theta \; d \phi, \qquad (\ell = a,b), \nonumber \\ &&
\end{eqnarray}
with  the
vector $\#q = q \, \hat{\#q}$, the scalars  $\kappa_\pm $ being the $q^2$ roots of $\mbox{det} \les
\underline{\underline{\mathbb{A}}}(\=U^{-1}\.\#q) \ris = 0$,
and the 6$\times$6 dyadics
\begin{equation}
\underline{\underline{\mathbb{A}}} (\#p)  = \les \begin{array}{cc}
\=\eps_{\,HCM} & \le \#p / \omega \ri \times \=I \vspace{2mm}  \\
-\le \#p / \omega \ri \times \=I  & \muo \, \=I
\end{array}
 \ris
\end{equation}
and
\begin{equation}
\underline{\underline{\mathbb{G}}}^{+} (\#p) =
\underline{\underline{\mathbb{A}}}^{-1} (\#p) - \lim_{| \#p | \to
\infty} \underline{\underline{\mathbb{A}}}^{-1} (\#p) .
\end{equation}}

\blue{
Generally, numerical methods are needed to evaluate the integrals on the right sides of Eqs.~\r{D_o} and \r{D_plus}, and  also
to extract the HCM permittivity dyadic
$\=\eps_{\,HCM}$  from the nonlinear dyadic equation~\r{Bruggeman_eqn} \c{Jacobi}.}

%\vspace{20mm}

%\noindent {\bf Tom G. Mackay}  is a Reader in the School of
%Mathematics at the University of Edinburgh, and also an Adjunct
%Professor in the Department of Engineering Science and Mechanics at
%Pennsylvania State University.
% His current research interests include homogenization, complex
%materials,  and sculptured thin films. He was elected
%a Fellow of SPIE in 2010.

\newpage

\newpage

\newpage

\newpage

\newpage

\newpage

\begin{thebibliography}{99}


\bibitem{BW}
M. Born  and E. Wolf, \emph{Principles of Optics}, 6th Edition,
Pergamon Press, Oxford, UK (1980).

%\bibitem{Khap}
%A. P. Khapalyuk, ``On the theory of circular optical axes,''
%\emph{Opt. Spectrosc. (USSR)} {\bf 12}, 52--54 (1962).

\bibitem{Voigt}
W. Voigt, ``On the behaviour of pleochroitic crystals along
directions in the neighbourhood of an optic axis,'' \emph{Phil. Mag.}
{\bf 4}, 90--97 (1902).

\bibitem{Panch}
S. Pancharatnam, ``The optical interference figures of amethystine
quartz~---~Part~II,''
\emph{Proc. Ind. Acad. Sci. A} {\bf 47}, 210--229 (1958).

\bibitem{Fedorov}
F. I. Fedorov and A. M. Goncharenko,  ``Propagation of light along the
circular optical axes of absorbing crystals," \emph{Opt. Spectrosc.
(USSR)} {\bf 14}, 51--53 (1963).

%\bibitem{Agranovich}
%V. M.  Agranovich  and  V. L. Ginzburg,   \emph{Crystal Optics with Spatial
%Dispersion, and Excitons}, Springer, Berlin, Germany (1984).

%\bibitem{Grech}
%B. N. Grechushnikov  and A. F. Konstantinova,  ``Crystal optics of
%absorbing and gyrotropic media," \emph{Comput. Math. Applic.} {\bf 16},
% 637--655 (1988).

\bibitem{Ranganath}
G. S. Ranganath, ``Optics of absorbing anisotropic media," \emph{Curr. Sci.} {\bf 67}, 231--237 (1994).

\bibitem{ML_PiO}
T. G. Mackay and A. Lakhtakia, ``Electromagnetic fields in linear bianisotropic
mediums," \emph{Prog. Optics} {\bf 51}, 121--209 (2008).

\bibitem{Lakh_helicoidal_bianisotropic_98}
A.  Lakhtakia, ``Anomalous axial propagation in helicoidal
bianisotropic media," \emph{Opt. Commun.} {\bf 157}, 193--201 (1998).

\bibitem{Berry}
M. V. Berry,  ``The optical singularities of bianisotropic crystals,"
 \emph{Proc. R. Soc.  A}
{\bf 461},  2071--2098 (2005).


\bibitem{ML03}
T. G. Mackay and A. Lakhtakia, ``Voigt wave propagation in biaxial
composite materials,''  \emph{J. Opt. A: Pure Appl. Opt. } {\bf 5},
91--95 (2003).


\bibitem{ML_WRM}
T. G. Mackay  and A. Lakhtakia, ``Correlation length facilitates Voigt
wave propagation," \emph{Waves  Random Media} {\bf 14}, L1--L11 (2004).

\bibitem{M2014_JNP}
T. G.  Mackay,
 ``On the sensitivity of directions that support Voigt wave propagation in infiltrated biaxial dielectric materials,"
  \emph{J.  Nanophotonics} {\bf 8},  083993 (2014).

\bibitem{M_JNP}
T. G.  Mackay,
 ``On extended homogenization formalisms for
nanocomposites,"
  \emph{J.  Nanophotonics} {\bf 2},  021850 (2008).

\bibitem{Ward} L.  Ward,  \emph{The Optical Constants
of Bulk Materials and Films}, 2nd edition, Institute of Physics,
Bristol, UK (2000).

\bibitem{EAB}
T. G.  Mackay    and  A. Lakhtakia,  \emph{Electromagnetic Anisotropy and
Bianisotropy: A Field Guide}, Word Scientific, Singapore (2010).


\bibitem{Arfken}
G. B. Arfken  and H. J. Weber, \emph{Mathematical Methods for
Physicists}, 4th Edition, Academic Press, London, UK   (1995).

\bibitem{GL01}
J. Gerardin  and  A. Lakhtakia, ``Conditions for Voigt wave propagation in
linear, homogeneous, dielectric mediums,'' \emph{Optik} {\bf 112},
493--495 (2001).

\bibitem{Boyd}
R. W. Boyd, \emph{Nonlinear Optics}, 2nd edition, Academic Press,
San Diego, CA, USA (2003).

\bibitem{LM07}
A. Lakhtakia and T. G. Mackay,
``Electrical control of the linear
optical properties of particulate composite materials,"
\emph{Proc. R. Soc. A}  {\bf 463}, 583--592 (2007).

\bibitem{Yariv_Yeh}
A. Yariv and  P. Yeh,   {\em Photonics: Optical Electronics in
Modern Communications}, 6th edition, Oxford University Press, New York,
NY, USA (2007).

\bibitem{L06_JEOS}
A.  Lakhtakia,
``Electrically tunable, ultra\-narrow\-band,
circular--polarization rejection filters with electro--optic
structurally chiral materials,"
\emph{ J. Eur. Opt. Soc.~--~Rapid
Pubs.} {\bf 1}, 06006 (2006).


\bibitem{Goncharenko}
A. V. Goncharenko,
``Generalizations of the Bruggeman equation and a concept of
shape--distributed particle composites,"
\emph{Phys. Rev. E} {\bf 68}, 041108 (2003).


\bibitem{WLM97}
W. S. Weiglhofer, A. Lakhtakia  and B. Michel,
``Maxwell Garnett and Bruggeman formalisms for a particulate composite with
bianisotropic host medium,''
{\em Microwave Opt. Technol. Lett.\/} {\bf 15}, 263--266 (1997); Erratum:
  {\bf 22},  221 (1999).

\bibitem{ZSB}
M. Zgonik, R. Schlesser, I. Biaggio, E. Volt, J. Tscherry, and P. G\"unter,
``Material constants of KNbO$_3$ relevant for electro-- and acousto--optics,"
\emph{J. Appl. Phys.} {\bf 74}, 1287--1297  (1993).% 1287--1297.

%\bibitem{MW_biax1}
%T. G. Mackay   and W. S. Weiglhofer, ``Homogenization of biaxial composite materials: Nondissipative dielectric properties,"
%\emph{Electromagnetics}  {\bf 21},  15--25  (2001).

\bibitem{MW_biax2}
T. G. Mackay   and W. S. Weiglhofer, ``Homogenization of biaxial
composite materials: dissipative anisotropic properties," \emph{J.
Opt. A: Pure Appl. Opt. } {\bf 2}, 426--432 (2000).


\bibitem{ML_JAP_2007}
T. G. Mackay and A. Lakhtakia, ``On scattering loss in electro--optic
 particulate composite materials,"
\emph{ J. Appl. Phys.} {\bf 101}, 083523
(2007).

\bibitem{M2011_JOPA}
 T. G. Mackay,
 ``Voigt waves in homogenized particulate composites
 based on   isotropic dielectric components," \emph{J.
Opt.} {\bf 13}, 105702  (2011).


\bibitem{Madou}
\blue{M. J. Madou, \emph{Fundamentals of Microfabrication: The Science of Miniaturization (2nd ed)}, CRC Press, Boca Raton, FL (2002).}

\bibitem{Kingon}
\blue{A. I. Kingon, S. H. Rou, M. S.
Ameen, T. M. Graettinger, K. Gifford, and O. Auciello, ``Deposition of electrooptic thin films,"  \emph{ Ceramic
Transactions} {\bf 14}, Electro-Optics and Non-linear Optic Materials, (Am.
Cer. Soc., Westerville, OH), 1990, pp.~179--196.}

\bibitem{Optical_switch}
T. S. El-Bawab, \emph{Optical Switching}, Springer, New York, NY, USA (2006).

\bibitem{M_WRM}
T. G.   Mackay,
 ``Depolarization volume and correlation length in the homogenization
of anisotropic dielectric composites," \emph{Waves  Random Media} {\bf
14}, 485--498 (2004); Erratum: \emph{Waves Random Complex Media} {\bf
16}, 85 (2006).

\bibitem{M97}
B. Michel,   ``A Fourier space approach to the pointwise singularity
of an anisotropic dielectric medium," \emph{Int. J. Appl. Electromagn.
Mech.} {\bf 8}, 219--227 (1997).

\bibitem{MW97}
B. Michel  and W. S. Weiglhofer,  ``Pointwise singularity of dyadic
Green function in a general bianisotropic medium," \emph{Arch.
Elekron. \"Ubertrag.} {\bf 51}, 219--223 (1997); Erratum: {\bf 52}, 31 (1998).


\bibitem{Jacobi}
B.  Michel,  A.   Lakhtakia and W. S.   Weiglhofer,
 ``Homogenization
of linear bianisotropic particulate composite media~---~Numerical
studies," \emph{Int. J. Appl.  Electromag.  Mech. } {\bf 9},
167--178 (1998); Erratum:  {\bf 10}, 537--538 (1999).


%\bibitem{Bohren}
%C. F. Bohren,  ``Do extended effective-medium formulas scale
%properly?,"
%\emph{J. Nanophotonics} {\bf 3}, 039501 (2009). %(2009)[doi:10.1117/1.3157171].


%\bibitem{M_Electromagnetics}
%T. G. Mackay,
% ``Effective constitutive parameters of linear nanocomposites in the long-wavelength regime,"
%  \emph{J.  Nanophotonics} {\bf 5},  051001 (2011).





\end{thebibliography}
\end{document}